\documentclass[screen, nonacm]{acmart}
\usepackage{enumitem}
\usepackage{hyperref}
\hypersetup{
    colorlinks=true,
    linkcolor=blue,
    filecolor=magenta,      
    urlcolor=cyan,
    }

\AtBeginDocument{%
  \providecommand\BibTeX{{%
    \normalfont B\kern-0.5em{\scshape i\kern-0.25em b}\kern-0.8em\TeX}}}

\begin{document}

\title[Transparency and Models of Conversational Search]{Understanding Mental Models of Generative Conversational Search and The Effect of Interface Transparency}

\author{Chadha Degachi}
\email{c.degachi@tudelft.nl}
\orcid{0000-0002-1850-689X}
\affiliation{%
  \institution{Delft University of Technology}
  \city{Delft}
  \country{The Netherlands}
}

\author{Samuel Kernan Freire}
\email{s.kernanfreire@tudelft.nl}
\orcid{0000-0001-8684-0585}
\affiliation{%
  \institution{Delft University of Technology}
  \city{Delft}
  \country{The Netherlands}
}

\author{Evangelos Niforatos}
\email{e.niforatos@tudelft.nl}
\orcid{0000-0002-0484-4214}
\affiliation{%
  \institution{Delft University of Technology}
  \city{Delft}
  \country{The Netherlands}
}

\author{Gerd Kortuem}
\email{g.w.kortuem@tudelft.nl}
\orcid{0000-0003-3500-0046}
\affiliation{%
  \institution{Delft University of Technology}
  \city{Delft}
  \country{The Netherlands}
}

\renewcommand{\shortauthors}{Degachi, et al.}

\begin{abstract}
   The experience and adoption of conversational search is tied to the accuracy and completeness of users' mental models---their internal frameworks for understanding and predicting system behaviour. Thus, understanding these models can reveal areas for design interventions. Transparency is one such intervention which can improve system interpretability and enable mental model alignment. While past research has explored mental models of search engines, those of generative conversational search remain underexplored, even while the popularity of these systems soars. To address this, we conducted a study with 16 participants, who performed 4 search tasks using 4 conversational interfaces of varying transparency levels. Our analysis revealed that most user mental models were too abstract to support users in explaining individual search instances. These results suggest that 1) mental models may pose a barrier to appropriate trust in conversational search, and 2) hybrid web-conversational search is a promising novel direction for future search interface design.
\end{abstract}

\ccsdesc[500]{Human-centered computing~Empirical studies in HCI}

\keywords{conversational search, information seeking behaviour, mental models, transparency}

\maketitle

\section{Introduction} 
Today, conversational search is a commercial reality~\cite{Janssen2024convoSearch} and the use of conversational agents (CAs) for search tasks such as finding, collating, and synthesising information has become accessible at an unprecedented scale\footnote{\href{https://web.archive.org/web/20240531084230/https://blogs.bing.com/search-quality-insights/february-2023/Building-the-New-Bing}{blogs.bing.com/search-quality-insights/february-2023/Building-the-New-Bing} ---last access \today}\footnote{\href{https://web.archive.org/web/20240610001101/https://openai.com/index/chatgpt/}{openai.com/index/chatgpt/} ---last access \today}\footnote{\href{https://web.archive.org/web/20240521051037/https://cloud.google.com/blog/products/ai-machine-learning/vertex-ai-io-announcements}{cloud.google.com/blog/products/ai-machine-learning/vertex-ai-io-announcements} ---last access \today}. 
This growth has been driven by the development of generative conversational agents (genCAs), which are 
are neural-net-based conversational systems built on models 
trained over extremely large corpora (large language models (LLMs))~\cite{Radford2018gpt1}. 
The nature of these models sidesteps the issue of “intent recognition” that plagued previous state-of-the-art CAs (e.g., Rasa\footnote{\href{https://rasa.com/docs/rasa/changelog/}{rasa.com/docs/rasa/changelog/} ---last access \today}), enabling better user experience (UX) in user-CA interactions~\cite{Freire2024llmvsrasa} and increasing conversational search popularity.

While the potential of user interaction in conversational search has been much discussed in literature~\cite{Zamani2023convoinfo, Zhang2018convosearch, Radlinski2017SearchCAFrame, Azzopardi2018CASearchInter}, understanding of CA search interactions, especially genCAs, remains limited. 
Emerging studies explore a diverging range of genCA search UX elements (e.g., search bias ~\cite{Sharma2024echogenSearch}, response explanations~\cite{lajewska2024exCS}, and over-reliance on search~\cite{Spatharioti2023llmsearch}), but underlying user perceptions and expectations remain under-explored~\cite{Datta2023whymms}. Where user expectations are not met, designers and developers risk poor long-term adoption, acceptance, and appropriate trust~\cite{Luger2016expecationsvoicedagetns,Lee2020CAmentalmodels, deVisser2020longtrustcalib,Ososky2013modelsHRT}. To investigate these perceptions and expectations, we seek to understand user mental models of genCAs.

Mental models are internal representations that individuals use to understand and interpret the world around them~\cite{norman2014some}, and evolve through interaction and training, enabling users to predict system behaviour~\cite{staggers1993mentalmodels, carroll1988mentalmodels, Phillips2011mentalmodels}. When mental models are more accurate, user satisfaction with the system improves~\cite{Kulesza2012mmagents} as does the appropriateness of trust~\cite{deVisser2020longtrustcalib,Ososky2013modelsHRT} and performance on search tasks~\cite{Kulesza2015explanations, Muramatsu2001transparentq}.
Mental models form through interaction with the system as facilitated by system affordances, such that users rely on “interaction clues”~\cite{Katzeff1990databaseMmodels} and system feedback to iterate upon these models until they become consistent~\cite{Katzeff1990databaseMmodels, Norman1988everydaythings}. 

Therefore, mental models become more aligned with actual system capabilities through interaction, if interaction with the system is sufficiently transparent. In fact, ~\citet{Luger2016expecationsvoicedagetns} pinpoint poor transparency in voiced digital assistants as a key reason behind users' inability to improve their mental models over time, contributing to low acceptance, and creating a gulf between user expectations and system reality. Nonetheless, many vectors for operationalising transparency have been explored within conversational and search interfaces. ~\citet{Muramatsu2001transparentq} show that communicating query transformations in web engine search improves user mental models of the system. 
Further, information source attribution has emerged as a vein of implementing conversational search transparency~\cite{Bohnet_2023GenCASource,Liu2023genCASearch, lajewska2024exCS}, where ~\citet{lajewska2024exCS} specifically, also explore source attribution in combination with confidence scores and system limitations in response explanations.

Inspired by decades of work exploring mental models of, and transparency in, information retrieval tools~\cite{Mlilo201110Mmodels, Zhang2008undergradMmodels, Katzeff1990databaseMmodels, Khoo2012searchmentalmodels, Thomas2020searchMM, Muramatsu2001transparentq, Bohnet_2023GenCASource, Liu2023genCASearch, lajewska2024exCS}, we propose two contributions to HCI 1) an explication of user mental models of generative conversational search, and 2) an investigation of the effect of interface transparency on mental models, user expectations, and search satisfaction. 
To that end, we pose the following research questions: 
\begin{enumerate}
    \item \textbf{RQ1}\label{rq:rq1}: What mental models of genCA search do users have?
    \item \textbf{RQ2}\label{rq:rq2}: What effect does the transparency of genCA search interfaces have on mental model, user expectations, and search satisfaction?
\end{enumerate}    

As surfacing latent mental models can be difficult, we design a mixed-method interview and think-aloud study~\cite{Zhang2008undergradMmodels, Zhang2024mmriskllms, Luger2016expecationsvoicedagetns,Katzeff1990databaseMmodels} involving 16 participants. Participants were given four search tasks and four CA interfaces. Interfaces used zero, one, two, or three transparency vectors (e.g., source attribution) chosen based on past research into search and CA transparency~\cite{Muramatsu2001transparentq, lajewska2024exCS, Liu2023genCASearch, Bohnet_2023GenCASource}. 

Our results show that most participants' understanding was too abstract to interpret individual search instances, with many mental models being inaccurate, incomplete, or both. Participants could, therefore, hold contradictory notions about genCAs, for example, conceptualising information sources as both unstructured corpus and structured databases simultaneously. 
Amid this uncertainty, participants were very concerned with both real and imagined system limitations, and attempted to compensate for these failures by creating hybrid conversational-web search workflows. 
Our results also suggest that while exposing agent function through transparency vectors stabilised expectation violation, it drove down satisfaction with the system. Moreover, transparency vectors improved system interpretability only when mental models were more complete. Based on these findings, we reflect on misaligned user-system trust in this, and past work. We also contribute a series of potential future research and design interventions for improved mental model alignment through greater interpretability, and thus more trustworthy and easier-to-adopt conversational search tools.


\section{Related Work}
\label{sec:rw}
To position this work in the literature, we 
review past research into users' experiences of conversational search, mental models of information retrieval tools and conversational agents, and the role of transparency in this context. 

\subsection{HCI in Conversational Search}
Conversational search systems are an emerging vein of interactive information retrieval~\cite{Zamani2023convoinfo} encompassing single and multi-turn natural language search interactions in "chat" interfaces alike~\cite{Liao2020convosearch}. Conversational search has been used for traditional information retrieval and recommendation tasks, as well as exploring modes of information retrieval unique to conversational agents, such as narrative search~\cite{Sadiri2024narrativesearch} and dialogue~\cite{Kiesel2021argumentsearch}. HCI studies of conversational search have emerged to predict user search satisfaction~\cite{Ghosh2023economicsconvosearch}, map conversational strategies
~\cite{Papenmeier2022convoStrategies, Xing2020voicedsearchpattern}, evaluate the usability or acceptability of a search agent~\cite{Arnold2020searchusefulness, Jung2019newsearchuse, Zelch2024adsearchaccept}, and investigate bias or inaccuracies in information retrieval~\cite{Sharma2024echogenSearch, Spatharioti2023llmsearch}. Of these studies, three explicitly relate to generative conversational search. ~\citet{Sharma2024echogenSearch}'s work on the effect of LLM-powered search on bias in information seeking,
~\citet{lajewska2024exCS}'s work on designing explanations for conversational search, and ~\citet{Spatharioti2023llmsearch}'s work to document search behaviour and overreliance in LLM-powered search. These studies have found a trend of overreliance on LLM-powered search tools, wherein users had difficulties noticing non-factual information in search responses~\cite{Spatharioti2023llmsearch} and explanations~\cite{lajewska2024exCS} without further highlighting in the interface. Moreover, ~\citet{Sharma2024echogenSearch} find that LLM-powered search biased results exploration more heavily than web engine search. 
Largely, we still lack an understanding of user perceptions and conceptualizations of conversational search, 
leading us to the study of mental models.

\subsection{Mental Models of Information Retrieval Tools and Conversational Agents}
Mental models are the frameworks \textit{``people have of […] the things with which they interact. People form mental models through experience, training, and instruction. The mental model of a device is formed largely by interpreting its perceived actions…''}- ~\citet[p. 17]{Norman1988everydaythings}. \citet{Zhang2008undergradMmodels} explicate mental models in the context of information retrieval by reporting on user perceptions of system components, functions, attributes, as well as their feelings about the system.
Mental models have been used to understand use of web search engines and databases for the past thirty years. Studies highlighted mismatches between user understanding and actual system behaviour, usability issues, and common attributes associated with the system~\cite{Zhang2008undergradMmodels, Katzeff1990databaseMmodels, Mlilo201110Mmodels, Khoo2012searchmentalmodels, Thomas2020searchMM}. They note that many users have a very simple understanding of how search works, and did not engage deeply with what search engines do. At the same time, research showed that user satisfaction with systems improves with mental model accuracy, as does search task performance~\cite{Kulesza2015explanations, Muramatsu2001transparentq}, and appropriateness of trust 
~\cite{deVisser2020longtrustcalib, Ososky2013modelsHRT, Mehrotra2023apptrustreview}, i.e., the extent to which that trust is aligned with trustworthiness~\cite{Jorge2021trusttrustworthiness,LeeSee2004trust, deVisser2020longtrustcalib}. 
Mental models which have been aligned with reality through training or system communication can calibrate expectations of, and trust in, systems to be more appropriate~\cite{johnson2021impact, Khastgir2018knowcalibration, bauhs1994knowing} thus moderating mis- and dis-use~\cite{parasuraman1997misuse, lee1994trust}. 
In conversational agents, \citet{Grimes2021expectationviolation} used mental models to quantify the extent to which user expectations moderated perceived competence of the agent and engagement with it. Similarly, researchers look at mental models of, and metaphors for, voice agents to explain the disuse of these tools over time ~\cite{Luger2016expecationsvoicedagetns,Lee2020CAmentalmodels}, and  understand user perceptions of  them~\cite{Desai2022alexametas}. ~\citet{Zhang2024mmriskllms} report on mental models of LLM-based CAs in relation to user perceptions of privacy risks in CAs, finding these models to be inaccurate and incomplete. We expand on this work by further analysing mental models of conversational search.

Eliciting mental models, due to their abstract nature, can be difficult. Often, researchers have combined multiple streams of input data to adequately represent user understanding of the system. Combinations of interviews~\cite{Zhang2008undergradMmodels, Zhang2024mmriskllms, Luger2016expecationsvoicedagetns}, think-aloud protocols~\cite{Zhang2008undergradMmodels, Katzeff1990databaseMmodels}, and illustration~\cite{Mlilo201110Mmodels, Zhang2008undergradMmodels, Zhang2024mmriskllms} 
have been used in the information retrieval and conversational agents domains. Research has also explored mental models of CAs through expectation violation, measured by self-report instruments~\cite{Grimes2021expectationviolation}. In this study, we have also adopted multiple data input streams; interviews, think-aloud, and expectation violation self-report.

\subsection{Aligning Mental Models Through Transparency}
\label{subsec:rwtransaprency}
The role of system transparency in supporting mental model formation is well established~\cite{Katzeff1990databaseMmodels, Norman1988everydaythings, Hoffman2023xaipsych}.  In fact, ~\citet{Radlinski2017SearchCAFrame} propose “System Revealment” — the system's revelation of its capabilities — as a key property of conversational search. 
The lack of transparency in conversational interfaces and its detrimental effects on mental models has similarly been explored. ~\citet{Luger2016expecationsvoicedagetns} and ~\citet{Motta2022CAmmodels} look to opacity and, specifically, the inability of the system to communicate its limitations and capabilities to users as a key failing of voiced digital assistants which has lead to poor mental models of these tools.

Though training and instruction have been used to align mental models with reality~\cite{johnson2021impact, Khastgir2018knowcalibration, bauhs1994knowing}, ~\citet{Zhang2008undergradMmodels} note that students formed mental models of web engines based on system cues and feedback over explicit instruction. ~\citet{Motta2022CAmmodels} also found that CA design professionals favoured subtle information dissemination through interaction over manuals.
So what vectors of implementing transparency within search interaction exist?
~\citet{Muramatsu2001transparentq} show that displaying how search queries are transformed 
aids in improving mental models of search. ~\citet{Spatharioti2023llmsearch} show that highlighting low and high confidence text segments can reduce overreliance on generated search responses. 
Further, information source attribution has emerged as a vein of implementing conversational search transparency~\cite{Bohnet_2023GenCASource,Liu2023genCASearch, lajewska2024exCS}. ~\citet{lajewska2024exCS} also  explored transparency through attribution, as well as confidence scores and communication of limitations in explanations, observing that perceived response usefulness improved, though explanations required more cognitive effort on the user's part. In fact, the field of explainable AI offers many vectors for implementing transparency within conversational interfaces, including increased and contextual user guidance~\cite{Motta2023transCA}, and user-review based recommendation explanations~\cite{Hernandez2021CAreviewXAI}. Outwith the conversational domain, other modes of explanation are also common, such as counterfactual explanations~\cite{Hoffman2023xaipsych} and confidence scores~\cite{Zhang2020xai, Hoffman2019xai}. Inspired by these works, we focus on three textual transparency vectors (1) Attribution, (2) Query Transformation, and (3) Confidence Signals which reflect common transparency vectors in information retrieval, as well as vectors designed specifically to intervene on mental model alignment (outwith conversational search).

\section{Methodology}
\label{sec:method}
This study asks 1) \textbf{RQ\ref{rq:rq1}}: What mental models of genCA search do users have? and 2) \textbf{RQ\ref{rq:rq2}}: What effect does the transparency of genCA search interfaces have on mental models, user expectations, and search satisfaction?

To address these questions, we designed a within-subject mixed-methods (interview, think-aloud, and self-report survey) experiment with 16 participants. Based on the vectors for implementing transparency outlined in Section \ref{subsec:rwtransaprency}, we built four interfaces graded 
from Baseline (no transparency vectors) to Most-Transparent (three vectors). Participants completed one search task per interface. 
A pilot study with two participants, not included in the final study, was conducted to assist us in finalizing the design of our conditions and tasks.

\subsection{Implementing the Conversational Search Interface Conditions}
\label{subsec:imp_details}
This work uses the term “conversational search” to refer to a search agent that supports a multi-turn (online or offline) information retrieval interaction using natural language in a dedicated conversational interface. To observe user interaction with such a system, 
we build four retrieval augmented generation (RAG), LLM-powered, prototypes. 
By embedding relevant context retrieved through vector search in prompts, RAG allows agents to improve the accuracy of genCA search and reduce incorrect results, making it better suited to information retrieval~\cite{shuster2021rag}. 

\begin{figure}
    \centering
    \includegraphics[width=0.8\linewidth]{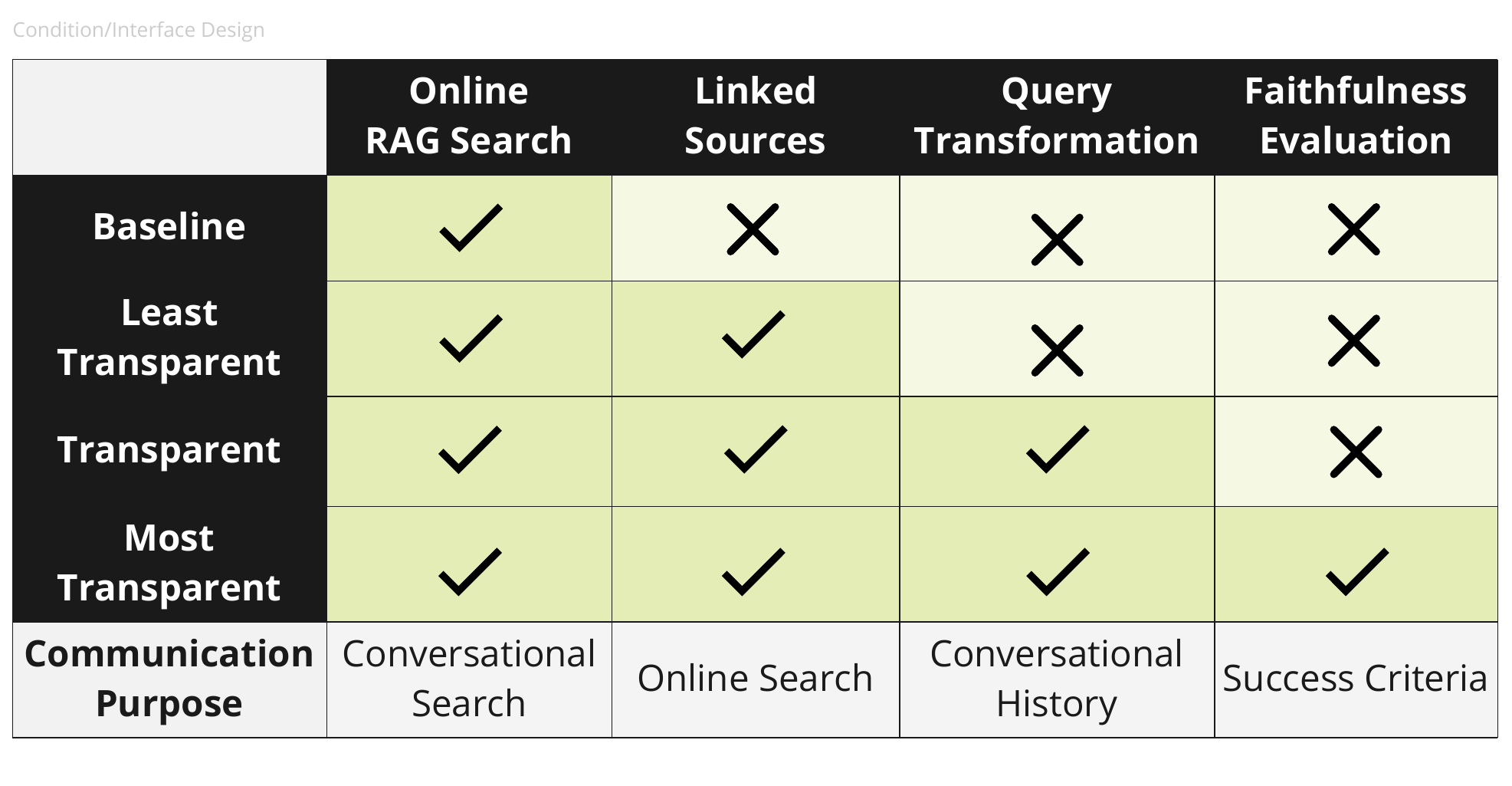}
    \caption{Summary of Conversational Agent Abilities and Purpose}
    \Description{Summary of Transparency Conditions. A 5 x 5 matrix with 4 conditions in 4 rows, and 4 agent features in 4 columns. Row 5 describes the purpose of vector per column. The table compares four transparency conditions: Baseline, Least Transparent, Transparent, and Most Transparent. Crosses and checks indicate which features are available in what condition, where the Most Transparent conditions has all features. Figure shows transparency levels in interface designs related to search functions.}
    \label{fig:ca_summary}
\end{figure}

Our four agents were built on \textit{gpt-4-0125-preview}\footnote{\href{https://web.archive.org/web/20240609034112/https://platform.openai.com/docs/models/gpt-4-turbo-and-gpt-4}{platform.openai.com/docs/models/gpt-4-turbo-and-gpt-4} ---last access \today} for response generations and the default \textit{OpenAI Embedding} model\footnote{\href{https://web.archive.org/web/20240527220522/https://platform.openai.com/docs/guides/embeddings}{platform.openai.com/docs/guides/embeddings} ---last access \today} for vectorizing documents. The agents also used the \textit{llmsearch} library\footnote{\href{https://web.archive.org/web/20240506081702/https://github.com/bdambrosio/llmsearch}{github.com/bdambrosio/llmsearch} ---last access \today}, the \textit{Google Search API}\footnote{\href{https://web.archive.org/web/20240602061755/https://developers.google.com/custom-search/docs/paid_element}{developers.google.com/custom-search/docs/paid\_element} ---last access \today}, and \textit{ZenRows}\footnote{\href{https://www.zenrows.com/}{zenrows.com} ---last access \today}. The interfaces were built on \textit{Gradio}\footnote{\href{https://www.gradio.app/}{gradio.app} ---last access \today}.  
The four agents' abilities are summarised in Figure \ref{fig:ca_summary}.

\begin{figure}
    \centering
    \includegraphics[width=0.8\linewidth]{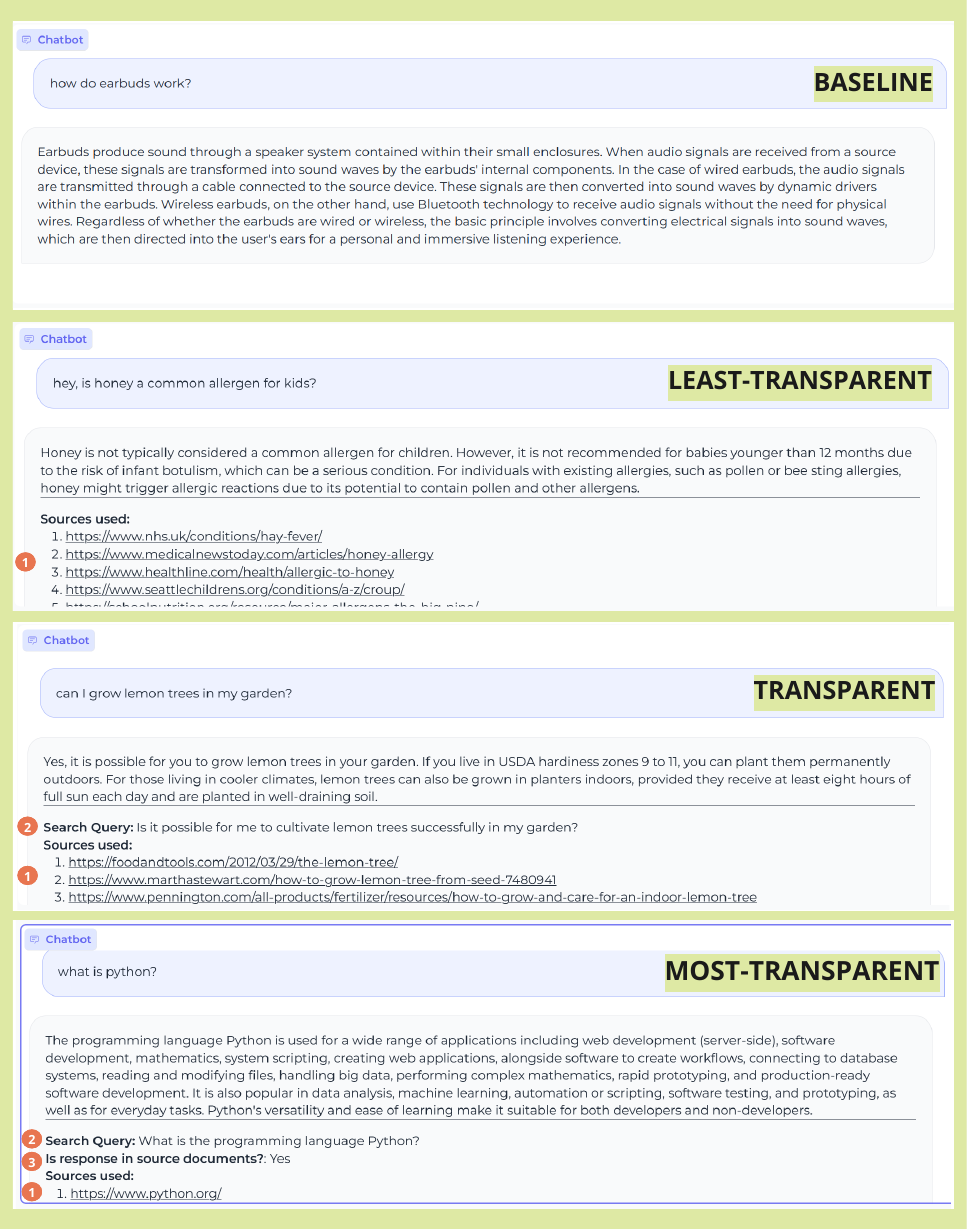}
    \caption{Baseline, Least Transparent, Transparent, and Most Transparent CA Interfaces}
    \Description{Four Conversational Search Interface Screenshots. Four subfigures are presented, in each subfigure two text bubbles, a question and a response, are shown. In subfigure 1, the response shows a block of text. In subfigure 2, the response shows a block of text followed by the header `Sources used:' and a numbered list of links. In subfigure 3, the response shows a block of text followed by the header `Search Query' with a line of text and the header `Sources used:' and a numbered list of links. In subfigure 4, the response shows a block of text followed by the header `Search Query' with a line of text. The next header reads `Is the response in source documents? Yes'. The last header reads `Sources used:' and a numbered list of links.}
    \label{fig:ca_interfaces}
\end{figure}

In the \textbf{baseline} condition, we presented only the text of search results in message bubbles. These bubbles were similar to the popular generative chat agent, ChatGPT, at the time of this study. The interface can be seen in Figure \ref{fig:ca_interfaces}, Sub-Figure 1.

In the \textbf{least-transparent} condition, we presented search results such that sources were included at the end of the message bubble after a line break. This format was similar to Bing Chat (or Copilot) at the time of this study. This common transparency vector was also used by ~\citet{lajewska2024exCS} in their conversational search explanation design. We choose this vector as a method of communicating “online search” functionality to users. The interface can be seen in Figure \ref{fig:ca_interfaces}, Sub-Figure 2.

In the \textbf{transparent condition}, sources and query transformations were included at the end of the agent's response after a line break. The transformed search query was obtained using an \textit{OpenAI} function, which uses the user input and chat history to create a more optimal query. Query transformations were studied by ~\citet{Muramatsu2001transparentq} for improving mental models of web engines. We chose query transformations as a method of communicating the use of conversation history in response generation to users. The interface can be seen in Figure \ref{fig:ca_interfaces}, Sub-Figure 3.

Lastly, in the \textbf{most-transparent condition}, sources, query transformations, and faithfulness were included at the end of the agent's response. Faithfulness was calculated using the \textit{llamaindex} evaluation metric, which “measure[s] if the response from a query engine matches any source nodes.\footnote{\href{https://docs.llamaindex.ai/en/stable/examples/evaluation/faithfulness_eval/}{docs.llamaindex.ai/en/stable/examples/evaluation/faithfulness\_eval} ---last access \today}.” Faithfulness is a popular method of assessing RAG performance~\cite{Es2023ragas, Ru2024ragchecker}, which we use as an analogue to the also popular confidence scores in decision support systems and the work of ~\citet{lajewska2024exCS}. 
We chose response evaluation to communicating agent “success criteria” to users. As faithfulness would be an obscure concept to most users, we design a natural language “Yes/No” flag with the label “Is response in source documents?” based on a simplified definition of faithfulness. The interface can be seen in Figure \ref{fig:ca_interfaces}, Sub-Figure 4.

\subsection{Participants}
We recruited 16 users from the research team's personal network using snowball and convenience sampling methods. As we aimed to collect extensive data from each participant for qualitative analysis, the sample size was deemed adequate for this goal. We selected participants who have been using generative conversational agents for search regularly in the past 12 months, as we targeted participants with pre-existing experience with conversational search. 
All participants had used ChatGPT, but participants also reported using Gemini ($n=7$), 
Copilot ($n=4$), 
and Custom Servers ($n=4$) 
Further, $56.25\%$ of participants reported they used genCA search at least weekly, of which three participants mentioned using genCA search daily. 
Nonetheless, $68.75\%$ of participants used web engines more often than they used chat search. 
The mean self-reported technical knowledge (collected using a 2-item scale introduced by \citet{Ashktorab2019pastknow}) among participants was \textit{8.125} (of a maximum 10), with a standard deviation of \textit{2.363}. Of 16 participants, 14 reported their demographic details. See Table \ref{tab:participants} for a summary of their characteristics.

\begin{table}[h]
\centering
\begin{tabular}{|l|l|}
    \hline
    Characteristics              & Distribution 
    \\ \hline
    Gender & 7 
    Female, 7 
    Male \\ \hline
    Age    & 1 
    < 25 years, 11 
    25-34 years, 2 
    > 34 years     
    \\ \hline
    Occupation or Area of Study &  7 Design, 6 
    Computer Science, 1 Other    
    \\ \hline
    Highest Level of Education  & 13
    Master's Degree,  1 
    Doctorate \\ \hline
    \end{tabular}
\caption{Summary of Participants' Characteristics}
\label{tab:participants}
\end{table}

\subsection{Materials \& Measures}
We combine semi-structured interviews, think-aloud,  conversation logs, and self-reported user experience metrics to capture a representation of user mental models as well 
the effect of transparency on mental models, user expectations, and search satisfaction. 

\begin{description}
    \item[Expectation Violation] We focus on capturing attributes related to the system, based on \citet{Zhang2008undergradMmodels}'s description of mental models. The attributes we choose are: comfort, humanness, skill, thoughtfulness, politeness, responsiveness, and engagement. These attributes were outlined as key aspects of social agents by \citet{Holtgraves2007socialagents} and later used (partially) by \citet{Grimes2021expectationviolation} to understand expectation violation. We captured expectation violation using the same 7-point Likert scale items twice, before and after interacting with the prototype, (Before: “I expect it is risky to interact with the chatbot”, After: “It is risky to interact with the chatbot”), in the same manner as~\citet{Grimes2021expectationviolation}. Participants were informed that they would interact with ``a chatbot to find information'' before rating their pre-task expectations, to capture general expectations for conversational search tools.  
    \item[Search Satisfaction] Measured by 
    a single item 7-point scale per search task. Satisfaction is a popular metric for assessing the usability of information retrieval tools
    ~\cite{liu2018searchsat}, it is associated with improved mental models~\cite{Kulesza2012mmagents}. We apply this measure here to identify possible `failures' in user-system interaction that may reveal gaps in user mental models.
    \item[Query Repairs] Captured in conversation logs. ~\citet{Kim2024queryrepair} show that some categories of follow-up queries in conversational search can be seen as query repairs and can signal user dissatisfaction. Thus, they may reflect points of ``failure'' in user-system interaction, and can reveal gaps in user mental models of the system.
\end{description}

Participants were asked to articulate their actions during a search using a concurrent, interactive, `Think Aloud' protocol~\cite{Boren2000thinkaloud,OBrien2023thinkaloud}. A semi-structured interview guide, based on the protocol developed by ~\citet{Zhang2008undergradMmodels} and our desire to probe at perceived system interpretability, was used to open and wrap up the study. The interview guide covered questions about system function, purpose, and components, as well as user understanding of the system. 

\subsection{Tasks}
Tasks were designed to require participants to locate more than one fact as to be sufficiently complex to engage participants. Tasks were written in the style of a personal scenario to encourage participants to extract the most relevant information from the text and formulate their own queries. They were presented using image cards to prevent “copy-pasting” task text. See Appendix \ref{app:task} for an example task. Tasks were related to 1) Recipe Substitution, 2) Art Movement Exploration, 3) E-Mail Service Recommendation, and 4) Mushroom Identification. We chose to focus on information retrieval tasks over other search tasks such as recommendations or travel planning to allow for an exploratory interaction where users would be interested in information factuality. Tasks were not time-limited. 

\subsection{Procedure}
\subsubsection{Research Design}
\begin{figure}
    \centering
    \includegraphics[width=0.5\linewidth]{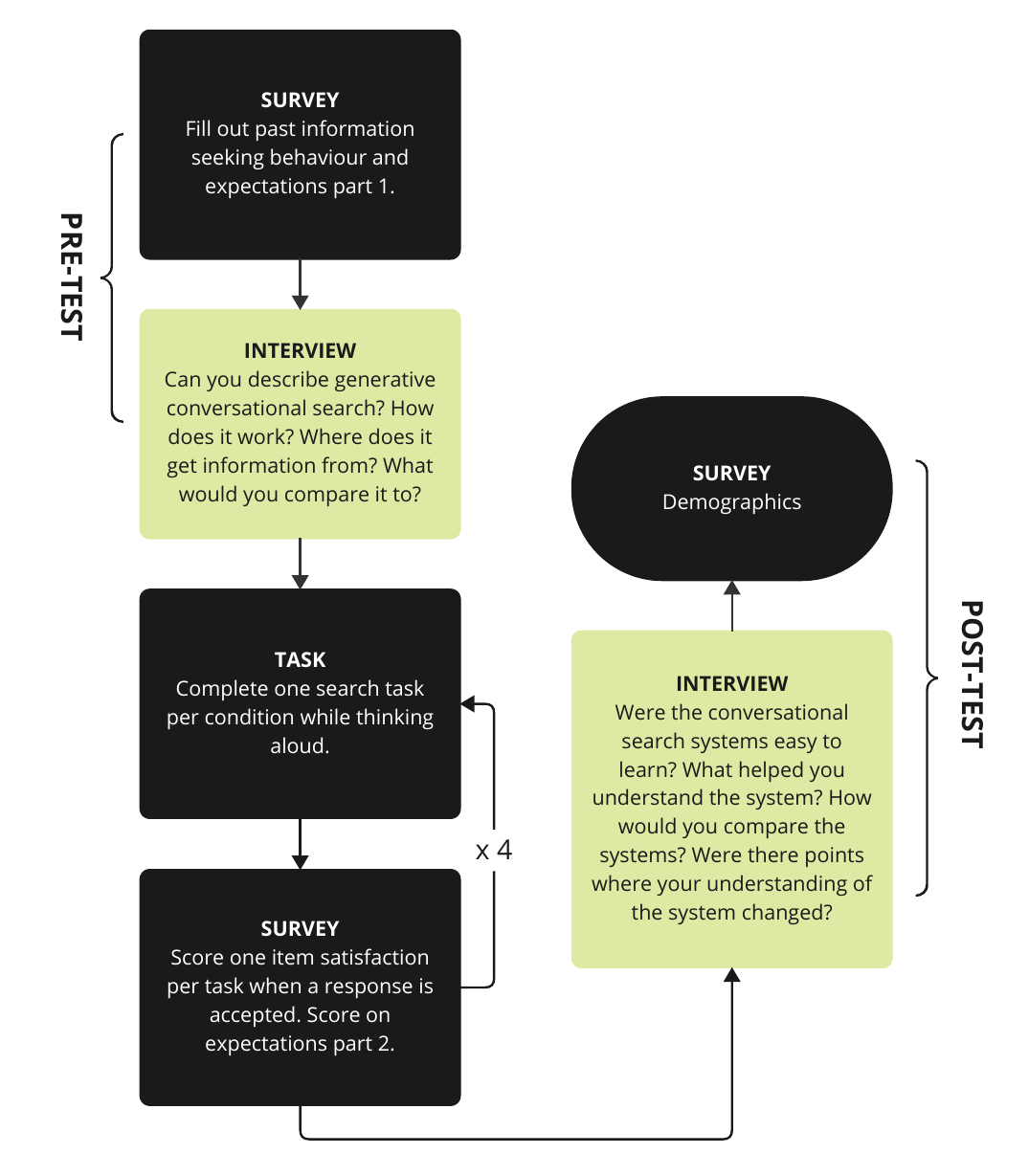}
    \caption{Study Procedure Flowchart}
    \Description{Study procedure flowchart outlines pre-test and post-test processes. The pre-test involves a survey on past information-seeking behavior and expectations, followed by an interview about generative conversational search. Participants then complete search tasks while thinking-aloud. The post-test includes a demographics survey and an interview discussing the ease of learning the system.}
    \label{fig:procedure_flow}
\end{figure}

The research design is seen in Figure \ref{fig:procedure_flow}. Interviews were scheduled using Calendly\footnote{\href{https://calendly.com/}{calendly.com} ---last access \today}~\cite{Harrison2022calendly} and Microsoft Teams\footnote{\href{https://www.microsoft.com/en-us/microsoft-teams/group-chat-software}{microsoft.com/en-us/microsoft-teams/group-chat-software} ---last access \today}. Participants were asked to join the online, recorded call with one of the research team members. 
They were then sent a link to the survey, in which they answered questions about their experience with search, as well as their expectations for the upcoming interaction. Once completed, the researcher conducted a 15-minute semi-structured interview with the participant, before sending them a link to the baseline agent and asking them to share their screen. 

Participants completed four untimed search tasks, one per condition. Tasks avoided overlap with common knowledge topics or likely participant expertise. Though participants were always exposed to conditions in the same order (from baseline to most-transparent), tasks appeared in a randomised order. Participants were allowed to combine the study genCA search interface with other web engines and genCA search tools to complete their task, as our observations of users during the pilot suggested this would be the most natural search behaviour. Participants were asked to verbalize their thoughts throughout the search task. After each task, the participant scored the system on satisfaction and the post-task part of the expectations survey. Once the tasks were completed, screen sharing was turned off, and another 15-minute semi-structured interview was conducted. Finally, the participants completed the demographics' survey. Participants were compensated with a voucher for their contribution. 

We chose to randomise task order but not condition order, as we were concerned that the appearance of disappearing functionality would confuse participants, and wished to gradually introduce interventions to the interface. This procedure was approved by our institution's human research ethics committee at (no.\textbf{[withdrawn for review]}).

\subsubsection{Data Analysis}

To capture the nuance of mental models, thematic analysis~\cite{clarke2013qual_research, braun2021thematic} was conducted on 16 interview and think-aloud transcripts to identify patterns of meaning within the data. After data familiarization, four transcripts were divided equally between the first two authors to independently develop a codebook through open, selective, and thematic coding~\cite{Braun2021tacodebook}. A combination of deductive and inductive orientation to the data was adopted. The authors then aligned the two codebooks in discussion to create the initial codebook. The codebook was iterated upon as needed throughout the analysis process so that code and themes could be refined. The remaining 12 transcripts were divided between the first two authors, with the first author analysing 8 transcripts. Having
multiple people analysing the data allowed us to develop rich
insights into the data.

For the survey responses, we used descriptive statistics and clustering to understand the shape of the collected data. We choose to study the data thus, as we did not have predetermined hypotheses to confirm. Content analysis~\cite{Harwood2003contentmethod} was applied to conversation logs and video logs to systematically categorize the frequency of query repairs in the interaction. This approach allowed us to quantify qualitative data and gain an overview of conversational search dynamics. The first author performed the coding on all 16 × 4 (64) conversation and video logs. 

\textit{Statement of Positionality:} The lead researcher on this work is situated within the field of human-computer trust and trust calibration. Further, the second, third, and fourth authors in this work are embedded in the domain of LLM adoption. These perspectives naturally colour our orientation towards our participants and our data.


\section{Results}
\label{sec:results}
Thematic analysis produced $161$ codes over $1319$ quotations and informed Sections \ref{subsec:importmodel} and \ref{subsec:mms}. Section \ref{subsec:trans} further combined this insight with content analysis and exploratory statistical analysis. These sections reflect two broad categories: 1) General Mental Models of Conversational Search (Sections \ref{subsec:importmodel} and \ref{subsec:mms}), and 2) Mental Models as Affected by Transparency in Study Conditions specifically (Section \ref{subsec:trans}).

\subsection{Imported Mental Models (RQ1)}
\label{subsec:importmodel}

Some ($N=6$) participants compared genCA search to human interaction. When they did, they referenced interactions with entry-level assistants, human domain experts, coworkers, and salesmen. \textit{``Because I use ChatGPT and Copilot mostly for work, I'd say they're like a[n] entry level assistant, saving me a lot of time and energy — especially at the beginning of the project to search for information — but, because there are entry level, I have to double-check everything.'' [P16]}. They discussed genCA search being opaque in the way human-human relationships can also be opaque, requiring time and effort to assess trustworthiness, capabilities, and limitations. Moreover, they described the agents as having intention in a human sense, and specifically, the intent to “please” the user, which leads agents to “oversell” their abilities. More participants compared genCA search to other technologies, including digital assistants, rule-based chatbots, and text prediction (autocomplete) algorithms. \textit{``My metaphor that I use when I explain it to others is that it is sort of an extreme version of […] when messages started sort of suggesting typing, right? […] It just has more context.'' — [P8]}. In terms of their similarity to search engines, there was some disagreement between participants, with some arguing that genCAs could not be compared to web engines, while others saw genCAs as “more powerful” web engines. 

In summary, we can see a wide array of both technical and human models being referenced by participants. In relation to RQ\ref{rq:rq1}, these results suggest that participants have a broad range of expectations from genCA search that do not converge.

\subsection{Mental Models of Conversational Search (RQ1)}
\label{subsec:mms}
\begin{figure*}
    \centering
    \includegraphics[width=\linewidth]{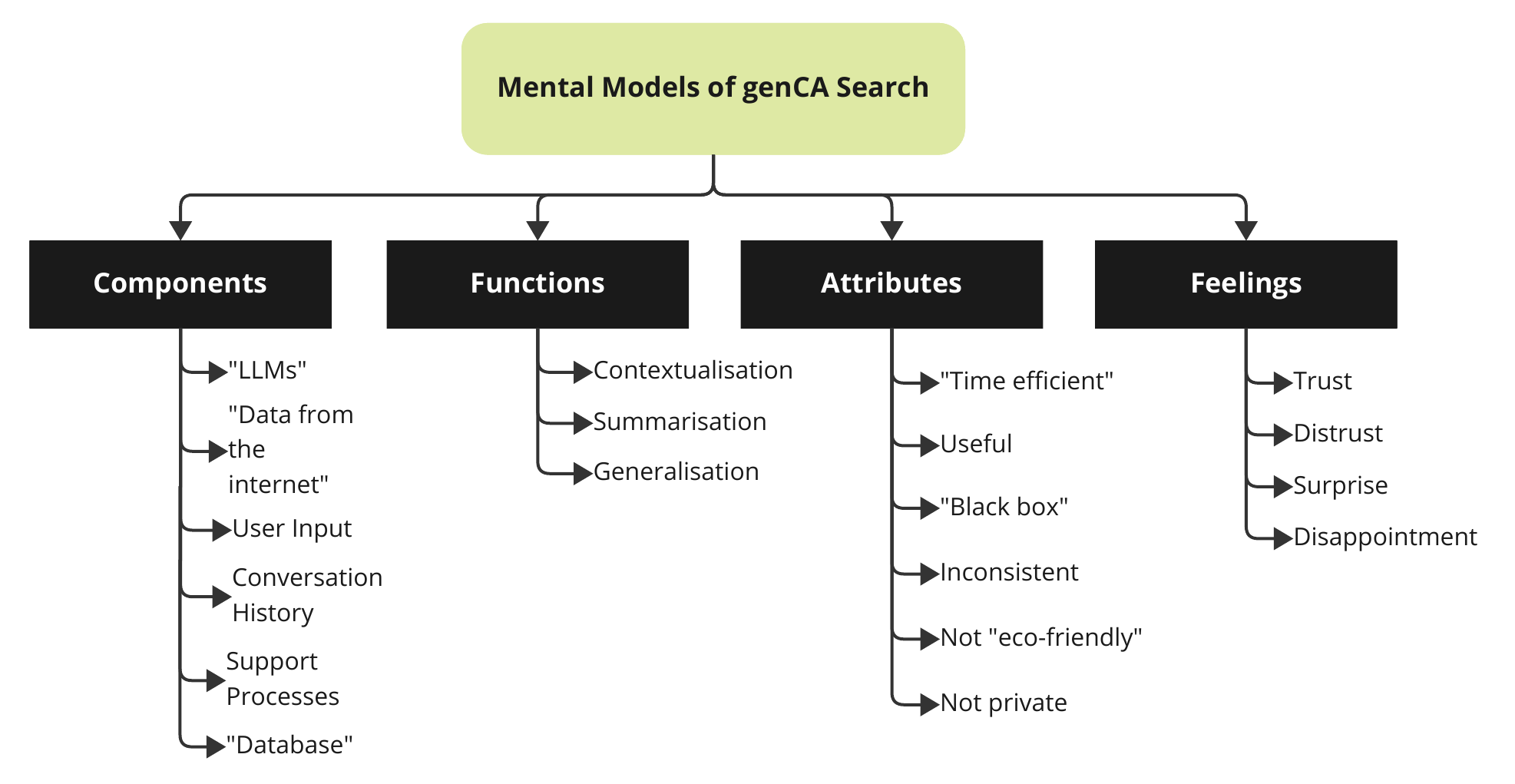}
    \caption{Components, Functions, Attributes, and Feelings of, and around, Conversational Search as Reported by Participants.}
    \Description{Diagram outlining mental models of genCA search. Models are devided into components, functions, attributes, and feelings. Components included LLMs, internet data, user input, conversation history, support processes, and databases. Functions were contextualization, summarization, and generalization. Attributes included time efficient, useful, black-box, inconsistent, not eco-friendly, and not private. Feelings associated were trust, distrust, surprise, and disappointment.}
    \label{fig:mm_results}
\end{figure*}

Figure \ref{fig:mm_results} shows a comprehensive overview of our results. We organise our findings along the framework for mental models of information retrieval used by \citet{Zhang2008undergradMmodels} wherein models are comprised of 1) Components, 2) Functions, 3) Attributes, and 4) Feelings.

\subsubsection{Components of Conversational Search}
Participants most consistently recognised and described three components of genCAs, namely 1) "Data from the internet", 2) "LLMs", and 3) some user input. 

The data corpus was described as the information source within genCAs. Often participants described this corpus as collected from “the internet”. When they were more specific, participants referred to web pages, blogs, and sometimes books, emails, and user posts. One participant mentioned data sources that they believed could not be in the training data, such as academic articles, as the companies creating the corpus would not have access to them. However, two other participants also referenced genCAs possibly illegally “scraping too much” data to train models. Few participants noted the use of their own conversation history in training new LLMs. 

Participants almost unanimously described genCAs as employing "LLMs" which use the "data from the internet". There was, however, little mention of neural nets, word-embedding, or transformers as part of the system. Overall, most participants were hesitant and unsure of their knowledge of this component beyond the use of LLMs, and described this component as a "black box". \textit{``I know that it's a statistical machine, let's say, so it's going to predict the most likely outcome to my request, and I don't actually know very much about how the answers are generated.'' [P12]}.

Though all participants recognised their query or input as part of the system, only some participants were aware that one could retrieve information from the genCA in a specified format and style using complex “prompts”, though few ($N=2$) used any complex prompting techniques such as examples to shape output. Formatting the output (list, table, images, etc.) was the most common prompt element. Few ($N=3$) participants knew of, and used, system prompts in their daily life. 

The less commonly recognised components of genCAs included support processes and the conversation history. Five participants ($N=5$) discussed the supporting components which enable the function of the genCA, such as function routing systems and moderation or guardrail processes. 
As for the conversation history, participants seemed to mostly be aware of this component implicitly, for example, by starting a new chat when they felt an agent is stuck, breaking down a search query so that the response to the initial query can be used to inform the next query, or formulating “incomplete” follow-up queries that rely on previous context. \textit{``I also tend to start a new chat. So because I assume that it is, and of course it is looking up all the other chats […] so that it would think from a fresh perspective.'' [P2]}.

Despite participants mostly identifying genCAs as using statistical models trained on a large unstructured corpus, there were still frequent references to databases in the participants' utterances. This was especially prominent when participants spoke about an individual search instance. Though many ($N=11$) participants were unclear on how the data, LLM, and user input interacted to produce a response, and described the system as a “black box” some also often referred to a “database” which is “searched” or “filtered”. From the same participant: \textit{``I've heard that it calculates the possibility of the appearance of [the] next word based on its training data.''}, then: \textit{``Yeah, it takes the question and then finds some corresponding resources in this database and then try to assemble [a response].'' [P10]}.

\subsubsection{Functions of Conversational Search}

In their descriptions of their day-to-day usage of genCA search, 
genCA search was often described as especially suited to contextualisation information tasks. 
These contextualisation tasks were described as supported by genCAs ability to collate and summarize or “average” information from many sources. Some participants felt that these systems were most suited to tasks on which there was a substantial amount of online information, and on which there was some degree of consensus online. \textit{``Like if I do a task that's been commonly done before, and I know it's, it's like a lot in the data that it's being fed, then it performs better than if it doesn't.'' — [P5]}. In turn, this view is echoed in the limitations of genCAs many participants cited, such as specificity, localization, currency, and semantics of information. A few participants were aware that most genCA tools were ``offline'' at the time of this study, with no access to up-to-date information, rendering them unsuitable for tasks such as news search. Other tasks seen as unsuitable for genCA search included those that were too high risk, too complex, too simple, or relied on mathematical reasoning. In terms of limitations in relation to factuality, one participant was aware that querying a genCA with “why” questions would make agent responses more consistent, in a Chain-of-Thought technique. Similarly, three participant were aware that using “leading” prompts would increase the odds of the agents returning a positive response, regardless of factuality.


\subsubsection{Attributes of Conversational Search}

Participants thought of genCA search as being efficient, useful, capable, and helpful. However, they also describe genCA search as being less private, less ``eco-friendly'', more ``data-hungry'', and more ``black box''
in comparison to web engines. Reducing search time was a major motivation behind participants' adoption of genCA search, so it was often described as both a feature of, and a requirement for, interacting with the system. \textit{``I'm using it to be time efficient, so I don't want to think too much about forming a story [when writing prompts].'' — [P9]}. 

Participants perceived genCA search as inconsistent and often discussed how they managed this aspect of the system through various output verification behaviours including web search, trial and error, multi-agent comparison, and agent interrogation. The perceived extent to which genCA search quality was inconsistent varied between participants, some found these agents to be ``good enough in most cases'' [P6], some found them often disappointing, one also noted that the quality of responses was dependent on traffic on the system at the time. \textit{``There have been times when it has surprised me, […] if I'm using it somewhere in the night, I don't know if it is because of traffic that it has on the website […] there is a huge difference in the way it generates a response.'' — [P2]}. 

\subsubsection{Feelings around Conversational Search}

Participants largely did not want to be surprised by the agent's output and became suspicious of the agent's response when it did not conform with their assumptions, even when the agent was correct, and the topic was not one they were knowledgeable about. 
Nonetheless, some participants seemed pleasantly surprised by the use of conversation history in the agents' response. They recognized this use when the response text referred to information disclosed or collected in a previous turn. 
Though, some participants did become frustrated when the agent was ``stuck'' in the conversation history while the participant was trying to redirect or correct the conversation. Participants expressed disappointment in agent responses when they were too general, irrelevant, or incorrect, but moreover, they expressed disappointed in agents that referred participants to human experts or other sources without returning an answer in ``risky'' topics. Though some participants saw this refusal as the correct or safe response, there remains an expectation of at least constructive guidance from the system. \textit{``So I would say that it did a spectacular job in like not killing me, but on the other hand, it did not complete the task.'' — [P4]}. 

Participants brought up feelings of trust and distrust towards genCA search fairly equally, with distrust being stronger. High task risk and complexity were contributing factors to distrust: \textit{``I'm a bit sceptical because I'm really afraid that these mushrooms are poisonous, and I wouldn't trust GPT or AI because I know that the technology is not that advanced yet.
'' — [P3]}. However, participants also distrusted genCA search correctness in general: \textit{``I mean, the information provided by Google can also be wrong. There could be fake news written by humans also, but it's a little bit more trustworthy for me compared to ChatGPT.'' — [P16]}. Information source visibility was mentioned as an antecedent of trust by participants, as well the information sources themselves. 
\textit{``I would say that my trust in ChatGPT is a little bit less compared to Copilot because Copilot is trained on the code repository from GitHub.'' — [P16]}. Participants relied on web search to confirm or verify the agent response with more ``trustworthy'' sources. 

Overall, most participants had an abstract understanding of genCA search. Few participants had more complete models of the system capabilities and limitations, such as online search and agreement with leading prompts, though these functions were directly related to concerns participants had about system limitations, such as inconsistency. Though many accurately described the global mechanism of generative conversational search (some model is trained over a large data corpus to take in natural language queries and generate a response), few could use this knowledge to explain individual search instances.  Therefore, the level of mental model abstraction allowed participants to hold erroneous contradictory notions, e.g. viewing information sources of genCAs as both an unstructured corpus and structured databases. Many participants were hesitant in describing the function or role of LLMs in the system, and described them as “black boxes”. Amid this uncertainty, participants, across mental model completeness levels, were very concerned with real and imagined system limitations. Participants dedicated search time to compensate for these failures in veracity, traceability, specificity, and trustworthiness by creating hybrid web search workflows. 
Nonetheless, participants both described genCA search as efficient and also prescribed efficiency to the system in their expectations and interaction modes.

\subsection{Effect of Interface Transparency (RQ2)}
\label{subsec:trans}
To understand how transparency affected user mental models, user expectations, and search satisfaction, we look at participant verbalisation during the search task and the post-test interview, perform content analysis on follow-up queries, and conducted an exploratory analysis of survey data. 

\subsubsection{Effect of Transparency on Mental Models of Components and Functions}
\label{subsubsec:transmms}
Across the three transparency vectors used in this study, sources were the most strongly noticed. Participants reacted very positively to their inclusion in the agent response, and often interacted with the links to verify output or further explore the search topic. 
However, sources did not strongly convey “online retrieval” functionality to participants. Some participants mentioned having had experience with genCA search tools generating “fake” links or returning sources post-hoc (e.g. Gemini's “Verify” function), therefore in interacting with these sources they were not sure of their roles. Secondly, some participants ($N=3$) with richer mental models, e.g. aware of online search and agreement with leading prompts, of genCA search did make the connection from sources to online search, therefore it could be that other participants were not aware this could be a function of genCA search. 

Query transformations were also noticed by many participants, but were not used to understand the role of conversation history. Instead, participants found query transformations useful in continuing their search task in a web engine or replicating the agent's search behaviour. \textit{``So then this is basically the search request. Yeah, that's actually that's pretty nice because then I can go and verify this myself and see if the searches are special.'' — [P6]}. In two cases, the combination of query transformation with sources did afford “online retrieval” to a participant. \textit{``Maybe the combination of query and link now that I think of it, maybe it's used that query, and then [returned] some of the top hits from Google in terms of the links to provide to me. 
'' — [P14]}


Of our 16 participants, only one participant mentioned feeling less judicious with output verification near the end of the study, while other participants discussed becoming increasingly critical of responses, especially when a past response was unsatisfactory or unexpected. \textit{``I just saw this [link]. It is completely irrelevant because it is not talking about mushrooms at all. […] So now I'm a little bit sceptical. Now I want to look [into] all the websites [to see] if I got the information right. 
'' [P2]}. However, some participants did note that the increased interpretability of the system with each condition was desirable. \textit{``For me, the first one is kind of completely a black box and the last one is kind of telling me how it works, [so] I will have more trust [in] the last one over the first.'' — [P7]}.

\subsubsection{Transparency, Expectations, and Search Satisfaction}
\begin{figure}
    \centering
    \includegraphics[width=0.8\linewidth]{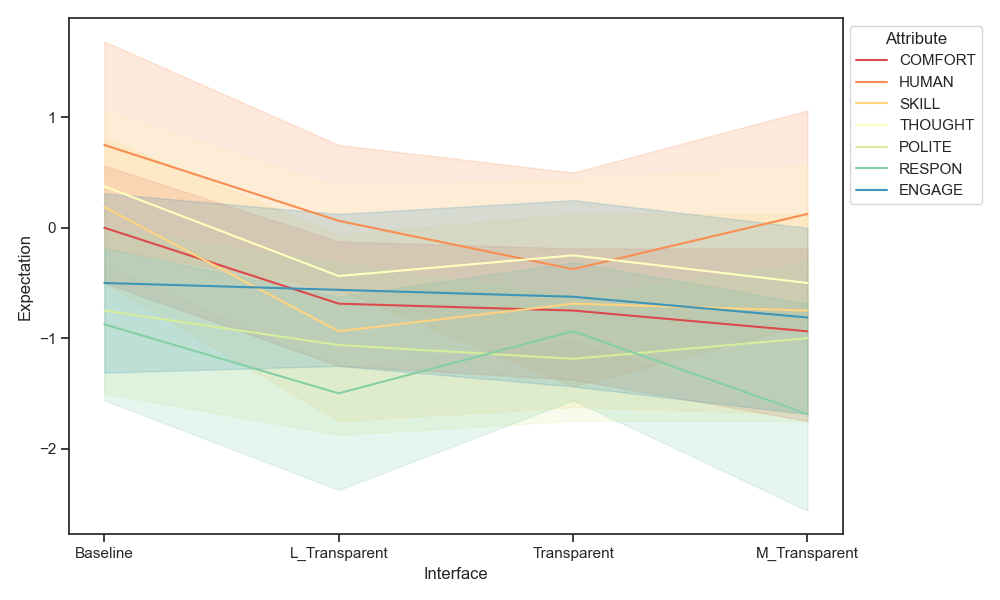}
    \caption{Expectation Violation (Delta of Pre-Test Expectation and Post-Test Rating) per Attribute per Interface}
    \Description{Expectation Violation Line Graph. A line graph depicting expectation violation (delta of pre-test expectation and post-test rating) for seven attributes (comfort, humanness, skill, thought, politeness, responsiveness, and engagement) across four interfaces (Baseline, L_Transparent, Transparent, M_Transparent). The graph shows a drop in expecatiations from Baseline to L_Transparent which then stabilizes.}
    \label{fig:expecation_viol}
\end{figure}

\begin{figure}
    \centering
    \includegraphics[width=0.5\linewidth]{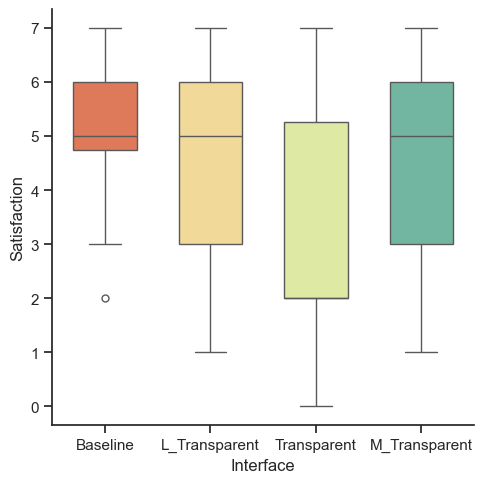}
    \caption{Search Satisfaction per Interface}
    \Description{Search Satisfaction box plot. A box plot comparing user satisfaction across four interfaces: Baseline, L_Transparent, Transparent, and M_Transparent. The plot shows variations in satisfaction levels with medians around 5 and downard trend with increased transparency vectors.}
    \label{fig:sat_per_interface}
\end{figure}

\begin{figure}
    \centering
    \includegraphics[width=0.8\linewidth]{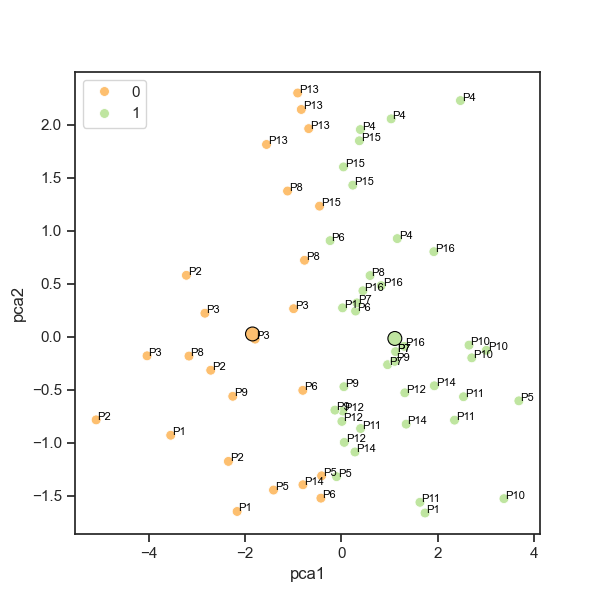}
    \caption{K-Means Clustering for Expectation Violation and Satisfaction Per Participant Per Interface. Cluster 0 represents participants with more negative expectation violation and search satisfaction, and Cluster 1 represents participants who were less so.}
    \Description{K-Means Clustering Scatter Plot on Expecations and Satisfaction. Axis labelled with two principal components, pca1 and pca2. Data points are split into two clusters, marked by color: orange (0) and green (1). Points are labeled P1 to P16 to indicate the associated participants. The plot shows Cluster 0 participants to be experience more negative expcatation violation and satisfaction than Cluster 1 participants.}
    \label{fig:expect_x_sat}
\end{figure}

Expectation violation (Figure \ref{fig:expecation_viol} and Appendix \ref{app:expect}) is most negative in the transition from the baseline to least-transparent interface, but is fairly stable from that point on (in all agent attributes expect responsiveness). 

Looking at survey data, we can see search satisfaction (Figure \ref{fig:sat_per_interface})\footnote{Where the median line is ``missing'' in a box plot, then the median line coincides with one of the quartile lines} remains relatively consistent across all interfaces, though with a downward trend with increased transparency vectors ($Mean_{satisfaction}=5.00, 4.60, 3.38, 4.33$ and $SD_{satisfaction}=1.41, 2.26, 2.36, 1.95$). As for query repairs, of 133 queries, 66 queries were new queries, 42 were follow-up queries, and 25 were query repairs. Four query repairs were related to a lack of specificity or localisation in the previous chat agent response. For example, \textit{``Can you be more specific to Italy and white mushrooms?”}. Fifteen query repairs related to a response from the agent that failed to answer the user's previous query; in these cases, the users reworded their previous query. 
Four repairs related to an intent misclassification in the previous turn, and two to the response being too long. Generally, the error recovery behaviour of participants involved negotiating with the limitations of genCA search by reformulating queries to be simpler (\textit{``Maybe I'm asking too much in one question.'' — [P13]}), more explicit (e.g., requesting output in a given format), or more detailed (\textit{``I'm gonna just paste [context from elsewhere] in at this point, when I'm a little bit lost like this. I kinda just dump info in the prompt and just see what happens.'' — [P6]}. Some participants also used follow-up queries to “interrogate” the agent's response when their trust was low, e.g., \textit{``Are you sure?''}. These findings suggest users were dissatisfied with CA responses in $18.80\%$ of cases, using ~\citet{Kim2024queryrepair}'s interpretation.

To understand how satisfaction and expectation violation interact, we cluster our participant's perceptions of both across all four interfaces using K-Means clustering on scaled and PCA (Principal Component Analysis) reduced data \cite{abdi2010principal}. Data loadings showed that violation of expected skill ($l=0.89$), thoughtfulness ($l=0.70$), and humanness ($l=0.65$) contributed most to principal component 1, while principal component 2 was characterised by responsiveness ($l=0.57$), comfort ($l=-0.52$), and politeness ($l=-0.51$). Therefore, participants in Cluster 0 were more disappointed in the agent's skill and responsiveness than others. We also find that participants consistently in Cluster 0 (\textit{P2, P3, P8, P13}) were also less familiar with components of genCA search such as online search, while Cluster 1 represented both abstract and complex mental models.

Returning to RQ\ref{rq:rq2}, the data suggests transparency might make users more aware of system limitations and errors, reducing the perceived skill of, and satisfaction with, the system, though it may stabilise expectation violation. Further, our results suggest that participants with more abstract mental models were also more disappointed with the system. 

\section{Discussion}
\label{sec:discuss}
Our findings suggest that most user mental models of genCAs are too abstract to support interpreting individual search instances (Section \ref{subsec:mms}), and that exposing search mechanisms through transparency vectors does not support the learning of the system when mental models are notably incomplete (Section \ref{subsec:trans}).

\subsection{Abstract Global Understanding but No Individual Interpretations}
Like the mental models of data handling in CAs reported on by ~\citet{Zhang2024mmriskllms}, our search mental models are also often incomplete and/or inaccurate. In both cases, these errors in understanding follow from users' too high-level and abstract understanding of conversational search, which in combination with an opaque interface left much room for self-contradicting system interpretations, especially at the local instance level. Moreover, we also see much concern about system inconsistency, as well as feelings of distrust and disappointment that may form barriers to effective genCA search utilization and adoption, as observed in voice assistants ~\cite{Luger2016expecationsvoicedagetns, Motta2022CAmmodels}. Indeed, users invested cognitive effort to incorporate genCA search into their workflows, 
creating personal rules around what tasks are appropriate for genCA search and 
building hybrid web-CA search paradigms
As for RQ\ref{rq:rq2}, our findings suggest that developing more interpretable genCA search systems that still ensure efficiency and user satisfaction will present a challenge for designers. Even while utilizing meta-information surrounding agent responses, and being concerned about agent inconsistency, participants still prioritised search task efficiency and reducing their own cognitive effort. This tension highlights a crucial and compelling design problem. 
Our users invested significant time compensating for the genCA's shortcomings by comparing search results across different platforms and search tools, while ~\citet{lajewska2024exCS} observe that their explanations required additional time and cognitive effort from users that may not correlate to the gain of using genCA search. 

How then to mitigate these knowledge gaps? Beyond interfaces, we know population-level communication avenues, such as advertising, influence user perceptions of technology~\cite{Kopitz2021alexaad}. Therefore, they may be used as design interventions to empower user understanding and anticipate mental model gaps before interaction starts. 
Further, in addition to trying to bridge knowledge gaps, search tools could mitigate some user burden by supporting more seamless hybrid web-CA search workflows, thus allowing users to 
compensate for real or perceived system limitations more efficiently. 


\subsection{Over-Trust and Under-Trust}
Overall, our findings are in step with previous investigations of mental models of information retrieval in web engines. Researchers have consistently found those user mental models in that domain to be relatively unsophisticated, and disengaged with what web engines do ~\cite{Zhang2008undergradMmodels, Thomas2020searchMM, Mlilo201110Mmodels, Khoo2012searchmentalmodels}. However, with too abstract mental models, user trust in genCA search is more likely to be misaligned, and through misalignment, increase the possibility of system mis- and dis- use~\cite{parasuraman1997misuse}. Indeed, where ~\citet{lajewska2024exCS, Spatharioti2023llmsearch} report overtrust and overreliance in their sample pool, we observe volatile trust and high distrust among ours. This distrust is evident in users' output verification behaviour, their feelings around genCA search and the system inconsistency they highlight. 

Moreover, our users' demonstrated unwillingness to be surprised by genCA search output, and the volatility of their trust in the system, can be paralleled to the observations of ~\citet{Sharma2024echogenSearch} in relation to bias in LLM-powered search. The authors highlight that users interacting with an agent biased towards them became more polarised in their opinion, but those interacting with one biased against them did not become more neutral in their position. ~\citet{Sharma2024echogenSearch}, in fact, note that their users paid more attention to text which agreed with them than that which did not, implying that users look for confirmation when they use genCA search.

\subsection{Implications for Design: Workflows of Future Search}
Our results have several actionable implications for the future design of conversational search, in order to improve system interpretability and support emergent user behaviour.
\begin{description}
    \item[Sources are Not Enough to Afford Online Retrieval Functionality:] Only users who had more complete mental models of genCAs prior to the study, e.g. knowledge of online search functionality, connected source attribution to online search (Section \ref{subsubsec:transmms}). This could be in part to generative agents generating sometimes fake links when requested, confusing users about the actual capabilities of the system. Sources were nevertheless very well received by participants, and used to support hybrid search workflows. 
    \item[Option to Turn Off Conversational History:] Memory is one of the five key properties of conversational search proposed by \citet{Radlinski2017SearchCAFrame}, however, based on our observations, it is not always a desirable property as users can feel stuck in a conversation, leading to users ``restarting'' conversations to obtain more satisfactory responses. Designers can better support this behaviour by allowing the use of conversation history to answer a query to be disabled when needed, or for previous conversational turns to open for modification (as ChatGPT does\footnote{\href{https://chat.openai.com/}{chat.openai.com}}). This feature would make users more aware of how their requests are processed and allow them more control over the system, as well as functioning as an implicit feedback mechanism for designers.  
    \item[Support Hybrid Workflows:] Users value web search for its ability to offer options and overviews in responses to allow users control over their search direction. Improving these aspects of conversational search, for example, through integrating conversation web browsing and supporting hybrid search workflows, could improve trust in the system while also compensating for information modalities which conversational search does not support, e.g. maps. 
\end{description}

\section{Future Work}
\label{sec:future}
Some aspects of conversational search and mental models lay beyond the scope of this study. 
For one, our work does not touch on mixed-initiative agents. 
Proactive agents have been posited as a great advantage of conversational search~\cite{Radlinski2017SearchCAFrame}, and future work should endeavour to investigate how proactivity impacts user understanding of the system. 
Further, based on the results of this work, we propose a series of hypotheses to be validated in future work in relation to genCA search interface design. 
\begin{description}
    \item[(H1) Out-of-Scope/Refer-to-An-Expert Answers Reduce Search Satisfaction:] Insights from this study suggest that users become frustrated when agents communicate the scope of their abilities by refusing to answer a query even as they are aware of the possible risk associated with a given query. Validating this insight would provide a starting point for understanding user information needs in conversational search. 
    \item[(H2) Out-of-Scope/Refer-to-An-Expert Answers Improve Trust:] Though frustrating to participants, clear knowledge scoping in high risk tasks was also desirable. Participants seemed to perceive this as an aspect of system trustworthiness. Validating this insight in combination with H1 can help outline the design space around information communication in genCA search.
    \item[(H3) Expectation Violation Moderates Trust More Negatively in genCAs Compared to Web:] With distrust and surprise so prominent in participants discussions of genCA search, trust in genCA search seems to be more volatile than user trust in web search, validating this insight would improve our understanding of the acceptance of novel technologies.
\end{description}

\section{Limitations}
\label{sec:limit}
This study has a few limitations, which we have worked to mitigate in our design. First, mental models are known to be difficult to articulate. Though we have mitigated this through our use of multiple elicitation techniques, participants may still hold beliefs that they were not aware of or could not express. Secondly, 
this study's ecological validity is both a strength and a potential detriment. Users phrased their queries naturally and the generative agent, as well as the Google Search API, returned results based on current accessibility and relevance. This means response quality and content varied between participants. While this enhances the realism of the findings, it also introduces variability that may affect consistency and reproducibility. Further, online search enabled conversational search interfaces were still emerging at the time of the study, mostly consisting of Bing Chat\footnote{\href{https://web.archive.org/web/20240531084230/https://blogs.bing.com/search-quality-insights/february-2023/Building-the-New-Bing}{blogs.bing.com/search-quality-insights/february-2023/Building-the-New-Bing} ---last access \today} and Perplexity\footnote{\href{https://www.perplexity.ai/}{perplexity.ai} ---last access \today}, but have become more commonplace since. Of course, the study was still conducted under conditions moderated by the researchers, and thus participant search behaviour may still differ from their in-the-field behaviour. Nonetheless, we offer a strong proxy to real-world interaction that reveals key insights into user mental models.
Lastly, our sample demographic (highly technical, highly educated, experienced young users), as well as sample size, limits the generalizability of our results but contributes to early insights into conversational search interaction. 

\section{Conclusions}
\label{sec:conclude}
In summary, we conducted a mixed-method interview and think-aloud study with 16 users to understand user mental models of genCA search and the effect of interface transparency on associated factors. We found most user mental models to be missing an understanding of key components and functions that would render individual search interactions more interpretable. Further, increased transparency did not aid learning when mental models were not already more accurate and complete. We recommend future research to further disentangle the relationship between search interface, mental models, user expectations, and search satisfaction.

\begin{acks}
This work is part of the NWO Our Smart Family Buddy project. Award number: KICH1.GZ01.20.016.
\end{acks}

\newpage

\bibliographystyle{ACM-Reference-Format}
\bibliography{main}

\newpage

\appendix
\section{Appendix: Task Presentation}
\label{app:task}
\begin{figure}
    \centering
    \includegraphics[width=0.5\linewidth]{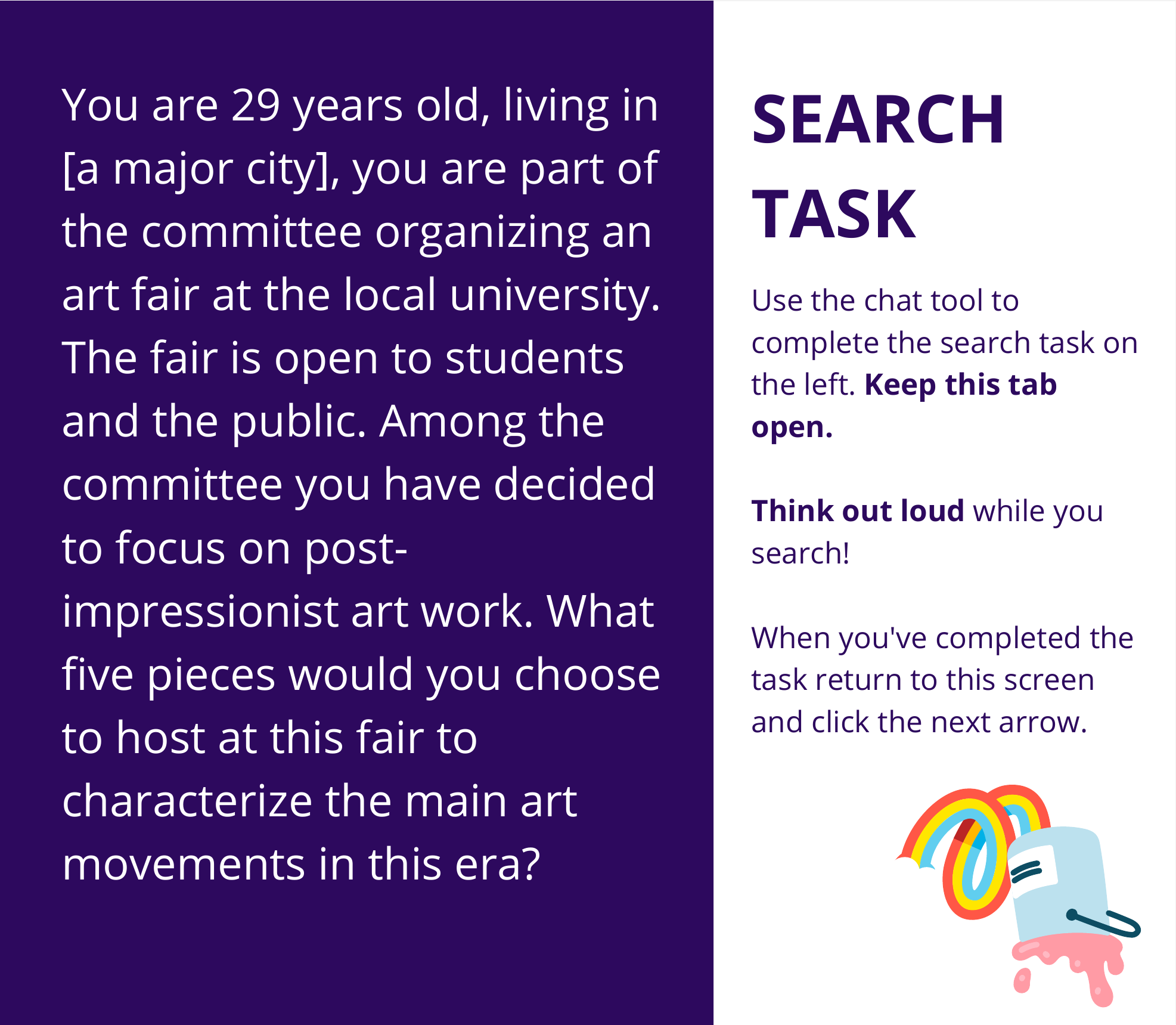}
    \caption{Example Search Task Card}
    \Description{Search Task Presentation. A card split into two parts. Part one is the search task text. Part two is the search instructions, along with two icons related to the task text.}
    \label{fig:example_task}
\end{figure}

Figure \ref{fig:example_task} shows a card with an example of the search tasks used in the study. 

\section{Appendix: Expectation Violation Per Attribute Per Interface}
\label{app:expect}
Table \ref{tab:expecation_violation} lists mean and standard deviation for expectation violation by attribute by interface.

\begin{table}[ht]
\centering
\begin{tabular}{|l|l|l|l|}
\hline
Interface      & Attribute      & Mean  & SD   \\ \hline
Baseline       & Comfort        & 0.00  & 1.15 \\ \hline
               & Engagement     & -0.50 & 1.83 \\ \hline
               & Humanness      & 0.75  & 1.84 \\ \hline
               & Politeness     & -0.75 & 1.48 \\ \hline
               & Responsiveness & -0.88 & 1.45 \\ \hline
               & Skill          & 0.19  & 1.38 \\ \hline
               & Thoughtfulness & 0.38  & 1.54 \\ \hline
Least-Transparent & Comfort        & -0.69 & 1.30 \\ \hline
               & Engagement     & -0.56 & 1.50 \\ \hline
               & Humanness      & 0.06  & 1.44 \\ \hline
               & Politeness     & -1.06 & 1.61 \\ \hline
               & Responsiveness & -1.50 & 1.83 \\ \hline
               & Skill          & -0.94 & 1.77 \\ \hline
               & Thoughtfulness & -0.44 & 1.67 \\ \hline
Transparent    & Comfort        & -0.75 & 1.24 \\ \hline
               & Engagement     & -0.63 & 1.82 \\ \hline
               & Humanness      & -0.38 & 2.03 \\ \hline
               & Politeness     & -1.19 & 1.17 \\ \hline
               & Responsiveness & -0.94 & 1.29 \\ \hline
               & Skill          & -0.69 & 1.85 \\ \hline
               & Thoughtfulness & -0.25 & 1.57 \\ \hline
Most-Transparent & Comfort        & -0.94 & 1.77 \\ \hline
               & Engagement     & -0.81 & 1.80 \\ \hline
               & Humanness      & 0.13  & 1.96 \\ \hline
               & Politeness     & -1.00 & 1.46 \\ \hline
               & Responsiveness & -1.69 & 1.99 \\ \hline
               & Skill          & -0.75 & 1.91 \\ \hline
               & Thoughtfulness & -0.50 & 2.28 \\ \hline
\end{tabular}
\caption{Mean And SD Of Expectation Violation (Delta Of Pre-Test Expectation And Post-Test Rating) Per Attribute Per Interface}
\label{tab:expecation_violation}
\end{table}

\end{document}